\documentclass[journal]{IEEEtran}
\usepackage{amsmath}
\usepackage[caption=false,font=normalsize,labelfont=sf,textfont=sf]{subfig}
\usepackage{dsfont}
\usepackage{amssymb,amsmath,amsfonts}
\usepackage{diagbox}
\usepackage{array}
\usepackage{multirow}
\usepackage{url}
\usepackage{bbm}
\usepackage{color}
\usepackage{xcolor}
\usepackage{tabularx}
\usepackage{cite}
\usepackage[normalem]{ulem}
\usepackage{booktabs}
\usepackage[caption=false,font=normalsize,labelfont=sf,textfont=sf]{subfig}
\usepackage{wrapfig}
\usepackage{graphicx}
\usepackage{textcomp}
\usepackage{stfloats}  
\usepackage{verbatim}
\usepackage{multirow}

\usepackage[normalem]{ulem}
\useunder{\uline}{\ul}{}
\usepackage[ruled,linesnumbered]{algorithm2e}
\def\BibTeX{{\rm B\kern-.05em{\sc i\kern-.025em b}\kern-.08em
T\kern-.1667em\lower.7ex\hbox{E}\kern-.125emX}}
\usepackage{balance}
\begin{document}
\title{MetaSTNet: Multimodal Meta-learning for Cellular Traffic Conformal Prediction}
\author{Hui~Ma, and Kai~Yang, \IEEEmembership{Senior Member, IEEE}
\thanks{This work was supported in part by National Natural Science Foundation of China under Grant 61771013.(Corresponding author: Kai Yang)}
\thanks{Hui Ma and Kai Yang are with the College of Electronic and Information Engineering, Tongji University, Shanghai 201804, China. (e-mail: kaiyang@tongji.edu.cn). }}

\markboth{Journal of \LaTeX\ Class Files,~Vol.~18, No.~9, September~2020}
{How to Use the IEEEtran \LaTeX \ Templates}

\maketitle

\begin{abstract}
Network traffic prediction techniques have attracted much attention since they are valuable for network congestion control and user experience improvement. While existing prediction techniques can achieve favorable performance when there is sufficient training data, it remains a great challenge to make accurate predictions when only a small amount of training data is available. To tackle this problem, we propose a deep learning model, entitled MetaSTNet, based on a multimodal meta-learning framework. It is an end-to-end network architecture that trains the model in a simulator and transfers the meta-knowledge to a real-world environment, which can quickly adapt and obtain accurate predictions on a new task with only a small amount of real-world training data. In addition, we further employ cross conformal prediction to assess the calibrated prediction intervals. Extensive experiments have been conducted on real-world datasets to illustrate the efficiency and effectiveness of MetaSTNet. 
\end{abstract}


\section{Introduction}
\IEEEPARstart{W}{ith} the rapid development of information and communication technologies, especially the emergence of Internet of Things and cloud computing, the network traffic pattern is becoming increasingly important\cite{9200470,DNA,PC2A,AIIOT}. According to the Cisco Annual Internet Report\cite{Cisco}, by 2023, there will be 29.3 billion network devices and 5.7 billion mobile subscribers worldwide, representing 71\% of the world's population. The explosion of mobile devices and applications has prompted the need for efficient network management and optimization. Network traffic prediction can alleviate network congestion and ensure stable network operations\cite{9293088}. Besides, it can reduce maintenance costs and improve quality of service for users\cite{9930825}. Therefore, it is crucial to improve the accuracy and confidence of cellular network traffic prediction.

To achieve accurate traffic prediction, deep learning approaches have received extensive attention\cite{Qiu_rnn,Kongitsg_2019,qin_2017,STaLSTMs,Wang2021a-3009159,ma2023cellular}. However, they rely on massive data and sufficient computing resources. In many real-world scenarios, it is time-consuming and expensive to acquire a large amount of data. When only a small amount of traffic data is available, deep learning models are prone to overfitting, resulting in poor generalization performance. Sim-to-real\cite {Feriani2021} is an effective way to address this problem as it can provide a large amount of data and reduce the cost and risk of training significantly\cite{Ge_2022}. Several studies\cite{Tobin2017,Ruiz2019,Shamsuddin2022} have demonstrated that using simulation data in a simulator can enhance performance in deep learning models. Inspired by this, we consider training a model with simulation data and transferring the learned knowledge to a real environment.

However, the model trained in a simulator may have degenerated performance in the real environment due to the reality gap. Although methods such as domain randomization\cite{Chen2022} and domain adaptation\cite{Zhao2020-survey} can somewhat alleviate this problem, they introduce a certain degree of randomness. Unlike the abovementioned methods, meta-learning can empower networks to learn to learn \cite{survey-meta-learning,Hospedales2021,Tian2022}. Motivated by this, we employ a meta-learning technique to solve this problem. We design a spatiotemporal prediction model based on a meta-learning framework, which enables the model obtained from the simulation environment to adapt quickly to a real environment.

In addition, cellular traffic data has complex spatiotemporal correlations\cite{b6,TWACNet,GCLSTM} and is influenced by multiple external factors. For instance, social activities and the spatial distribution of urban buildings are directly related to the generation of cellular traffic\cite{STCNet}. However, it is challenging to exploit multi-source heterogeneous auxiliary information to infer highly time-varying traffic patterns\cite{summaira2021recent}. Multimodal learning can comprehensively characterize traffic patterns by integrating complementary information captured from multimodal data. Thus, we consider using multimodal learning to learn a better feature representation by extracting and fusing spatiotemporal dependencies from multimodal data, which can help improve prediction accuracy.

Existing accurate estimation of network traffic has focused on point prediction. However, the predicted volumes are susceptible to noise and model inference errors \cite{Abdar2021}, resulting in a lack of trust in the predictions. Therefore, it is vital to assess the confidence of predicted volumes in this context. By understanding the inherent uncertainty in predictions, decision-makers can make more accurate risk assessments and decisions. As one of the uncertainty quantification approaches, conformal prediction has received much attention in quantifying the confidence of predictions. The fundamental advantage of conformal prediction over other techniques is that it does not require any specific assumptions on data distribution except for the i.i.d. (or exchangeable). Besides, it is a model-agnostic method that can be used with many machine learning models\cite{meta-cp}. For instance, cross conformal prediction is employed in \cite{Gupta2022} that divides the training data using $K$-fold cross validation with computationally efficient. However, it is unsuitable for time series prediction. On this basis, we design a data split strategy for time series and use it in cross conformal prediction to obtain prediction intervals.


To tackle the abovementioned challenges, we propose an end-to-end spatiotemporal model, namely MetaSTNet, for cellular traffic prediction. Specifically, MetaSTNet differs from existing cellular traffic prediction in three aspects. \textit{\textbf{First}}, we construct a simulation platform to simulate cellular network traffic data at the cellular level. Then, we train the prediction model with a bi-level optimization procedure\cite{ji2021bilevel}. In the inner-level optimization stage, we train the model with simulation data and obtain the meta-knowledge (i.e., initialized model parameters). In the outer-level optimization stage, the learned initialized parameters can be further optimized and the meta-knowledge learned from the simulation environment can be directly transferred and adapted quickly to a new task with a small amount of real-world traffic data. \textit{\textbf{In addition}}, we propose event-driven attention to capture the long-term temporal correlations of traffic and textual data. Also, we employ GCN and CNN to characterize the spatial dependencies of traffic and image data, respectively. Then, we design a spatiotemporal block (ST-block) to fuse temporal and spatial relationships. Furthermore, we design two paralleled encoder-decoder structures to capture the short-term (hourly) and periodic (daily) characteristics of multimodal data, respectively. \textit{\textbf{Finally}}, we employ a growing-window forward-validation scheme to split training traffic data and use it in cross conformal prediction when only a small amount of traffic data is available. Specifically, we divide the training data of the target task into $K$ disjoint subsets (folds). The $k, k \in\{1, ..., K-1\}$-th fold is used to fine tune initialized parameters, while the next $(k+1)$-th fold is adopted to calculate nonconformity scores and use them to obtain prediction intervals. 

The contributions of our study can be summarized as follows:
\begin{itemize}
\item \textbf{Cellular traffic prediction with simulation:} We propose a deep learning model with sim-to-real transfer techniques and leverage the approximate implicit differentiation based approach to solve the bi-level optimization problem in meta-learning. In fact, we are the first to employ simulation data for cellular traffic prediction when only a small number of real-world traffic data is available.

\item \textbf{Spatiotemporal patterns mining based on multimodal data:} We propose a novel spatiotemporal multimodal fusion approach for cellular traffic prediction. As far as we know, this is the first work to explore the fine-grained and multi-scale feature fusion method of multimodal data (i.e., traffic data, textual data, and image data) for cellular traffic prediction.

\item \textbf{Uncertainty estimation with cross conformal prediction:} we employ a growing-window forward-validation scheme for cross conformal prediction to quantify the confidence of predicted volumes by constructing valid prediction intervals. To the best of our knowledge, this work takes the first step to explore interval prediction in cellular network traffic and assess the confidence of predictions.
\end{itemize}

Extensive experiments have been carried out on three real-world datasets to evaluate the prediction accuracy of our proposed MetaSTNet. It is revealed that it exhibits strong empirical performance and outperforms state-of-the-art techniques.

The remainder of this paper is organized as follows. We provide a brief introduction of related work in Section II. Then we describe the problem formulation and the details of our proposed model in Section III. After that, we present detailed experimental setups in Section IV and analyze experimental results in Section V, respectively. Finally, we conclude our study in Section VI.

\section{Related Work}

In this section, we first investigate the characterization and modeling approaches for cellular traffic prediction (see Section II-A). Then, we discuss previous works on traffic prediction methods based on a meta-learning framework (see Section II-B). Finally, we survey a set of related work regarding conformal prediction methods (see Section II-C). For convenience, Table \ref{tab1} summarizes the techniques and associated abbreviations in our study. Besides, Table \ref{tab2} summarizes the latest related work and highlights their main characteristics.

\subsection{Cellular Traffic Prediction}
Cellular traffic forecasting techniques can be classified into statistical and machine learning approaches. Popular statistical prediction methods like ARIMA and ES have simple structures and can infer linear relationships among multiple variables. They work well with stationary time series but perform poorly when predicting turning points. Machine learning techniques are proposed to model the complex nonlinear characteristics of traffic data, which can be further classified into traditional machine learning and deep learning approaches. Several studies employ traditional machine learning methods such as RFR, SVR, and HMM\cite{aceto2021characterization} for network traffic prediction. However, they require manual feature extraction and are computationally inefficient when a large number of high-dimensional training samples are available.

\begin{table}[]
\renewcommand\arraystretch{1.25}
\centering
\caption{Acronyms techniques mentioned in our study}
\label{tab1}
\begin{tabular}{ll}
\hline
Acronyms & Technique \\ \hline
ARIMA & AutoRegressive Integrated Moving Average \\
CCP &   Cross Conformal Prediction  \\
ConvLSTM & Convolutional Long Short Term Memory\\
CNN & Convolutional Neural Network   \\
ES &  Exponential Smoothing  \\ 
FFN & FeedForward Network \\
GCN &  Graph Convolutional Network    \\
GRU &  Gated Recurrent Unit   \\ 
HMM & Hidden Markov Model \\
ICP &  Inductive Conformal Prediction   \\
LSTM & Long Short-Term Memory    \\ 
MLP & Multilayer Perceptron     \\ 
RFR & Random Forest Regressor \\ 
RNN & Recurrent Neural Network     \\ 
SVR & Support Vector Regression\\ 
\hline
\end{tabular}
\end{table}

Deep learning approaches have been widely used in network traffic prediction since they can handle a large amount of data and characterize volatile traffic patterns. The RNN-based methods are widely used to capture the long-term temporal dependencies of cellular traffic\cite{Qiu_rnn,Kongitsg_2019,qin_2017,STaLSTMs,Wang2021a-3009159}. Wang \textit{et al}.\cite{LSTM_GP} characterized periodic features with Fourier transform and modeled time-domain correlations with LSTM. However, they have difficulty in capturing the spatial correlations of cellular traffic. To characterize spatial dependencies\cite{SAE,STGCN-HO,GNN-D,attention}, Shen \textit{et al}.\cite{TWACNet} extracted the long-term temporal and spatial correlations with time-wise attention and CNN, respectively. Fang \textit{et al}. \cite{GCLSTM} adopted GCN and LSTM to extract the spatiotemporal dynamics of traffic data. In addition, several attempts have been made to model short-term, periodic, and trend patterns of traffic data\cite{DeepAuto}. In\cite{STACN}, three paralleled attention and convolution modules were designed to capture the spatiotemporal dependencies of hourly, daily, and weekly traffic data. Zhang \textit{et al}.\cite{DeseNet} characterized local and periodic traffic patterns using densely connected CNN. Liu \textit{et al}.\cite{ST-Tran} proposed a novel structure to capture short-term and periodic spatiotemporal dependencies. 

Furthermore, previous studies show that external information can improve the prediction performance of cellular traffic\cite{DeepAuto,STACN,Yaoweiran-sq-2021}. For instance, Bhorkar \textit{et al}. proposed DeepAuto\cite{DeepAuto}, which studied the impact of network configurations, holidays, and other auxiliary factors for cellular traffic prediction. Montieri \textit{et al}.\cite{montieri2021packet} used the window size of transmission control protocol to improve prediction performance. Jie \textit{et al}.\cite{DeepTP} explored points of interest, number of weeks, and population flows for traffic prediction. Zhang \textit{et al}.\cite{STCNet} analyzed cross-domain datasets, including the number of base stations, points of interest, and social activity levels, for cellular traffic prediction. 

\subsection{Meta Learning}
Meta-learning methods can be categorized into metric-based, model-based, and optimization-based approaches\cite{survey-meta-learning,Hospedales2021,Tian2022}. The metric-based techniques aim to learn a similarity score to determine the similarity between two samples. The model-based approaches aggregate information from all tasks into memory units (internal states) and then obtain predictions on a new task\cite{Pan2021,Fang2021-META-MSNet,Zhang2021-dmTP}. The optimization-based methods use multiple tasks to learn favorable initialized parameters of the prediction model so that it can converge in a few iterations on a new task\cite{Wang2022-MetaTTE,Li2021-STG-Meta,Yao2019-ST-NET,Zhang2020}.

Meta-learning has been widely used in time series prediction. In terms of one-dimensional time series forecasting, Zhang \textit{et al}.\cite{Zhang2021-dmTP} used LSTM as a meta-learner to develop a model-based meta-learning framework. Oreshkin \textit{et al}.\cite{meta-nbeats} designed N-BEATS, which can also be viewed as a meta-learning method. In terms of high-dimensional time series forecasting based on meta-learning, Fang \textit{et al}.\cite{Fang2021-META-MSNet} used MLP as a meta-learner to extract external information so as to construct the parameters of fully convolutional networks. Besides, a set of meta-learners are utilized to model different types of traffic\cite{19971829}. Orozco \textit{et al}.\cite{Orozco2020} proposed a probabilistic regression model that could learn information from multiple tasks and form a shared feature embedding. Moreover, the optimization-based meta-learning methods have been applied to urban traffic predictions\cite{Yao2019-ST-NET,Li2021-STG-Meta}. When the training data of the target city is insufficient, the spatiotemporal characteristics can be extracted from multiple cities in close proximity and used to improve prediction accuracy.

\subsection{Conformal Prediction}

\begin{table*}[]
\renewcommand\arraystretch{1.25}
\centering
\caption{a collection of prediction techniques in related work}
\label{tab2}

\begin{tabular}{cccccc}
\hline
Technique &  Spatiotemporal dependency & Time series observations & auxiliary data\\ \hline
Fourier transform$+$LSTM\cite{LSTM_GP} &  & Single  &   \\
DeseNet\cite{DeseNet}  & \checkmark & Multiple&   \\
ST-Tran\cite{ST-Tran}  & \checkmark & Multiple&    \\
DeepRTP\cite{Liu2019}  & \checkmark & Multiple&    \\
ST-DenNetFus\cite{ST-DenNetFus}   & \checkmark & Multiple & Unimodal  \\
TWACNet\cite{TWACNet}  & \checkmark & Multiple& Unimodal \\
STACN\cite{STACN} &  \checkmark & Multiple& Unimodal \\
DeepAuto\cite{DeepAuto}  & \checkmark & Multiple & Unimodal  \\
STCNet\cite{STCNet}  & \checkmark & Multiple & Unimodal  \\ \hline 

N-BEATS\cite{meta-nbeats} &   & Single &   &  \\
Meta-MSNet\cite{Fang2021-META-MSNet} & \checkmark & Multiple & Unimodal   \\
STG-Meta\cite{Li2021-STG-Meta} & \checkmark & Multiple &   Unimodal  \\
MetaST\cite{Yao2019-ST-NET} & \checkmark & Multiple &   Unimodal \\ 
dmTP\cite{Zhang2021-dmTP} &   & Single &  Unimodal \\
\hline

EnbPI\cite{EnbPI} &  & Single &     \\
CF-RNN\cite{CF-RNN} &   & Multiple &    \\
CQR\cite{CQR} &   & Single &   \\
EnCQR\cite{jensen2022ensemble} &   &Multiple &     \\
stableCP\cite{ndiaye2022stable} &   &Single &    \\
\hline
MetaSTNet &\checkmark &  Multiple & Multimodal  \\\hline
\end{tabular}
\end{table*}

Conformal prediction has been widely used in regression tasks\cite{Zeni2023,LVDPI,Toccaceli2022}. Angelopoulos \textit{et al}. \cite{Angelopoulos2021a} provided a recent survey on conformal prediction and \cite{Lei2018DistributionFreePI,barber2022conformal,paolo_2022} discussed the variants of conformal prediction, which reveal that full conformal prediction is computationally inefficient as its training time is much higher than other variants. In contrast, inductive conformal prediction\cite{CQR} has a light computational overhead since it divides training samples into two disjoint subsets, including a training set and a calibrating set to train the model and calculate nonconformity scores, respectively\cite{jensen2022ensemble}. Kamilė \textit{et al}.\cite{CF-RNN} applied inductive conformal prediction to high-dimensional time series forecasting. However, when training data is insufficient, a small training set may make it difficult to obtain favorable model parameters. In contrast, a small calibrating set may lead to a high variance of confidence since calibrating nonconformity scores becomes unreliable.

To balance the computational cost and prediction accuracy, jackknife+ and CCP are proposed\cite{barber2021predictive,Gupta2022}. Xu \textit{et al}.\cite{EnbPI} proposed ensemble estimators to construct prediction intervals for dynamic time series. Besides, Barber \textit{et al}. \cite{barber2021predictive} mentioned that jackknife+ can be reckoned as a special case of CCP when the size of each subset is equal to one. In CCP, a smaller $K$ means that we only need to fit the model $K$ times with the cost of slightly larger residuals. Gupta \textit{et al}. \cite{Gupta2022} adopted CCP to obtain prediction intervals for classification and regression tasks. So far, however, there has been little research about exploiting conformal prediction for spatiotemporal cellular traffic forecasting.

\section{Problem Formulation and Framework}
\subsection{Definitions and Problem Formulation}
\subsubsection{Definition 1}
We focus on multi-step ahead time series prediction. In our study, it is assumed that there are $N+1$ tasks, including $N$ auxiliary tasks and a target task. Besides, all tasks obey the same distribution, i.e., $\mathcal{T}_i \sim{P(\mathcal{T})},i\in\{1,...,N+1\}$. $\mathcal{T}_i$ is the $i$-th task and $P(\mathcal{T})$ denotes the distribution over all tasks. In each task $\mathcal{T}_i$, we use multimodal data with three modalities, including cellular traffic, textual, and image data. The traffic data can be expressed as $\boldsymbol{X}_{\mathcal{T}_i}^{\rm{tra}}=[\boldsymbol{x}_{\mathcal{T}_i,1}^{\rm{tra}},...,\boldsymbol{x}_{\mathcal{T}_i,T}^{\rm{tra}}]^\top \in \mathbb{R}^{T\times D}$, where $\boldsymbol{x}_{\mathcal{T}_i,t}^{\rm{tra}}$ denote the $t$-th element of $\boldsymbol{X}_{\mathcal{T}_i}^{\rm{tra}}$ and $T,D$ denote the length and dimension of traffic data. The external information extracted from textual data can be represented as $\boldsymbol{X}_{\mathcal{T}_i}^{\rm{txt}}=[\boldsymbol{x}_{\mathcal{T}_i,1}^{\rm{txt}},...,\boldsymbol{x}_{\mathcal{T}_i,T}^{\rm{txt}}]^\top \in \mathbb{R}^{T\times D_{\rm{txt}}}$, where $D_{\rm{txt}}$ represents the dimension of external data. The image data is denoted as $\boldsymbol{X}_{\mathcal{T}_i}^{\rm{img}}\in \mathbb{R}^{W\times H\times C}$ and $W, H, C$ represent the width, height, and channel of image data, respectively. Our goal is to predict traffic volume $\boldsymbol{y}_{T+h}\in \mathbb{R}^{D}$ at time $T+h$, where $h$ denotes the number of steps predicted in advance. For simplicity, we omit the subscript $T+h$ and use $\boldsymbol{y}$ to denote output data.

\subsubsection{Definition 2}
In our study, the dataset is composed of meta-training and meta-testing data. The meta-training data is associate with $N$ auxiliary tasks, i.e., $\mathcal{D}_{\rm{meta\_train}}=\{(\mathcal{D}_{\mathcal{T}_1}^{\rm{tr}},\mathcal{D}_{\mathcal{T}_1}^{\rm{val}}),...,(\mathcal{D}_{\mathcal{T}_N}^{\rm{tr}},\mathcal{D}_{\mathcal{T}_N}^{\rm{val}})\}$. 
In task $\mathcal{T}_i$, the multimodal data can be further divided into a training set (support set) $\mathcal{D}_{\mathcal{T}_i}^{\rm{tr}}$=$\{\boldsymbol{X}^{\rm{tra,tr}}_{\mathcal{T}_i},\boldsymbol{X}^{\rm{txt,tr}}_{\mathcal{T}_i},\boldsymbol{X}^{\rm{img}}_{\mathcal{T}_i},\boldsymbol{y}_{{\mathcal{T}_i}}^{\rm{tr}}\}$ and validating set (query set) $\mathcal{D}_{\mathcal{T}_i}^{\rm{val}}$=$\{\boldsymbol{X}^{\rm{tra,val}}_{\mathcal{T}_i},\boldsymbol{X}^{\rm{txt,val}}_{\mathcal{T}_i},\boldsymbol{X}^{\rm{img}}_{\mathcal{T}_i},\boldsymbol{y}_{\mathcal{T}_i}^{\rm{val}}\}$.

The meta-testing data is associated with the target task, which can be divided into training, calibrating, and testing sets, i.e., $\mathcal{D}_{\rm{meta\_test}}=\{\mathcal{D}^{\rm{tr}},\mathcal{D}^{\rm{cal}},\mathcal{D}^{\rm{ts}}\}$. For simplicity, we omit the subscript $\mathcal{T}_{N+1}$ when denoting meta-testing data. And they can be represented as $\mathcal{D}^{\rm{tr}}$=$\{\boldsymbol{X}^{\rm{tra,tr}},\boldsymbol{X}^{\rm{txt,tr}},\boldsymbol{X}^{\rm{img}},\boldsymbol{y}^{\rm{tr}}\}$, $\mathcal{D}^{\rm{cal}}$=$\{\boldsymbol{X}^{\rm{tra,cal}},\boldsymbol{X}^{\rm{txt,cal}},\boldsymbol{X}^{\rm{img}},\boldsymbol{y}^{\rm{cal}}\}$, $\mathcal{D}^{\rm{ts}}$=$\{\boldsymbol{X}^{\rm{tra,ts}},\boldsymbol{X}^{\rm{txt,ts}},\boldsymbol{X}^{\rm{img}},\boldsymbol{y}\}$, respectively. 
     
\begin{figure}[!t]
\centering   
\includegraphics[width=\columnwidth]{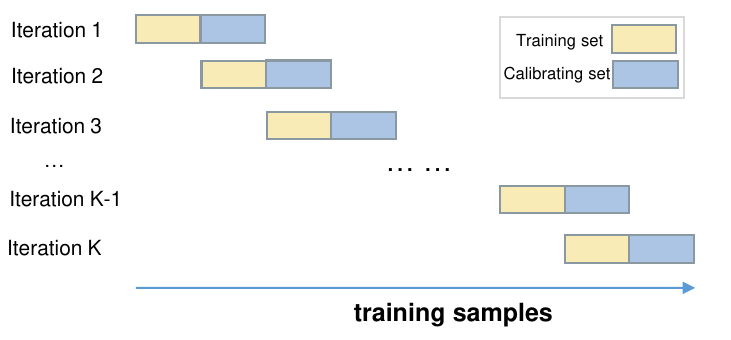}
\caption{Splitting training samples from the target task to generate training and calibrating sets.}
\label{data_split}
\end{figure}

As is shown in Fig. \ref{data_split}, training samples are divided into $K$ disjoint subsets (folds) according to the growing-window forward-validation scheme\cite{Schnaubelt2019comparison}. The $k,k\in\{1,...,K-1\}$-th fold (yellow color) represents a training set used to fine tune model parameters. And the next $(k+1)$-th fold (blue color) denotes a calibrating set used to obtain nonconformity scores. 

\subsubsection{Problem formulation}
In this work, we design a network traffic prediction model (MetaSTNet) based on a meta-learning framework, which is composed of meta-training and meta-testing stages.

The meta-training phase takes a bi-level optimization procedure\cite{ji2021bilevel}. We assume $\widetilde{\boldsymbol{\omega}}=(\boldsymbol{\omega}_1,...,\boldsymbol{\omega}_N)$, where $\boldsymbol{\omega}_i,i\in \{1,...,N\}$ denotes the task-specific parameters. In the inner-level optimization stage, the base learner of task $\mathcal{T}_i$ seeks $\boldsymbol{\omega}_i$ as the minimizer of its loss on training set $\mathcal{D}_{\mathcal{T}_i}^{\rm{tr}}$ in Eq. (\ref{Eq.3}). And $\widetilde{\boldsymbol{\omega}}^*$ can be optimized over all $N$ tasks in Eq. (\ref{Eq.2}). In the outer-level optimization stage, the meta-learner evaluates the minimizer $\boldsymbol{\omega}^*_i$ on validating set $\mathcal{D}^{\rm{val}}_{\mathcal{T}_i}$, and optimizes $\boldsymbol{\theta}$ over all $N$ tasks in Eq. (\ref{Eq.1}). To be specific, $\boldsymbol{\omega}_i$ is the parameters of the last linear layer of MetaSTNet while $\boldsymbol{\theta}$ corresponds to the parameters of the remaining layers\cite{Ji2020}. Moreover, $g(\boldsymbol{\theta},\widetilde{\boldsymbol{\omega}})$ represents the loss function of inner-level optimization stage and  $f(\boldsymbol{\theta},\widetilde{\boldsymbol{\omega}}^*)$ denotes the objective function of outer-level optimization stage.
\begin{equation}
\min _{\boldsymbol{\theta}} \Psi(\boldsymbol{\theta})= f(\boldsymbol{\theta},\widetilde{\boldsymbol{\omega}}^*)=\frac{1}{N} \sum_{i=1}^{N} \  \frac{1}{\left| \mathcal{D}_{\mathcal{T}_i}^{\rm{val}} \right|} \sum_{\xi \in \mathcal{D}_{\mathcal{T}_i}^{\rm{val}}} \mathcal{L}_i(\boldsymbol{\theta},\boldsymbol{\omega}^*_i;\xi), 
\label{Eq.1}   
\end{equation}
\begin{equation}
\label{Eq.2} 
\rm{s.t.} \ \widetilde{\boldsymbol{\omega}}^* = \arg \min _{\widetilde{\boldsymbol{\omega}}} g(\boldsymbol{\theta},\widetilde{\boldsymbol{\omega}})=\frac{1}{N} \sum_{i=1}^{N}\ \mathcal{L}_{\mathcal{S}_i}(\boldsymbol{\theta},\boldsymbol{\omega}_i),
\end{equation}
\begin{equation}
\label{Eq.3} 
\mathcal{L}_{\mathcal{S}_i}(\boldsymbol{\theta},\boldsymbol{\omega}_i) = \frac{1}{\left| \mathcal{D}_{\mathcal{T}_i}^{\rm{tr}} \right|} \sum_{\xi \in \mathcal{D}_{\mathcal{T}_i}^{\rm{tr}}} \mathcal{L}_i(\boldsymbol{\theta},\boldsymbol{\omega}_i; \xi).
\end{equation}

To solve the problem Eq. (\ref{Eq.1}), we employ the approximate implicit differentiation based approach as it is computationally and memory efficient\cite{ji2021bilevel}. In the inner loop, we can obtain $\omega_{i,j}^{P}$ for each meta-training task $\mathcal{T}_i$ using $P$ steps of gradient decent, where $\widetilde{\boldsymbol{\omega}}_j^{P}=(\boldsymbol{\omega}_{1,j}^{P},...,\boldsymbol{\omega}_{N,j}^{P})$ and $j\in \{1,...,J\}$ denotes the number of epochs. Then, we aim to estimate hypergradient $\hat{\nabla} \Psi (\boldsymbol{\theta}_j)$ in the outer-level optimization stage. To be specific, as shown in \cite{ji2021bilevel}, we first obtain $v^{Q}_j$ by solving $\nabla^2_{\widetilde{\boldsymbol{\omega}}}g(\boldsymbol{\theta}_j,\widetilde{\boldsymbol{\omega}}_j^P)v=\nabla_{\widetilde{\boldsymbol{\omega}}}f(\boldsymbol{\theta}_j,\widetilde{\boldsymbol{\omega}}_j^P)$ with $Q$ steps of conjugate gradient. Then, we can obtain the Jacobian-vector product $\nabla_{\boldsymbol{\theta}}\nabla_{\widetilde{\boldsymbol{\omega}}} g(\boldsymbol{\theta}_j,\widetilde{\boldsymbol{\omega}}_j^{P})v^{Q}_{j}$ with existing automatic differentiation packages. Finally, we can obtain hypergradient estimation with the following equation\cite{ji2021bilevel},
\begin{equation}
\widehat{\nabla}\Psi(\boldsymbol{\theta}_j)=\nabla_{\boldsymbol{\theta}}f(\boldsymbol{\theta}_j,\widetilde{\boldsymbol{\omega}}_j^{P})-\nabla_{\boldsymbol{\theta}}\nabla_{\widetilde{\boldsymbol{\omega}}} g(\boldsymbol{\theta}_j,\widetilde{\boldsymbol{\omega}}_j^{P})v^{Q}_{j}.
\label{Eq.4}
\end{equation}

In the meta-testing phase, we use the training set $\mathcal{D}^{\rm{tr}}$ to fine tune the model parameters, which can be denoted as $\boldsymbol{\theta}^{*}$. Then, we exploit the calibrating set $\mathcal{D}^{\rm{cal}}$ and yield estimate $\boldsymbol{\hat{y}}^{\rm{cal}}$,
\begin{equation}
\boldsymbol{\hat{y}}^{\rm{cal}}=f_{\boldsymbol{\theta} ^{*}}(\boldsymbol{X}^{\rm{tra,cal}},\boldsymbol{X}^{\rm{txt,cal}},\boldsymbol{X}^{\rm{img}}),
\end{equation}
where $\boldsymbol{\hat{y}}^{\rm{cal}}=[\hat{y}^{\rm{cal}}_1,...,\hat{y}^{\rm{cal}}_D]$ and $\hat{y}^{\rm{cal}}_i$ denotes the $i$-th element of $\boldsymbol{\hat{y}}^{\rm{cal}}$.

Suppose there are $L$ number of samples in $\mathcal{D}^{\rm{cal}}$. Then, the nonconformity score $R_i, i\in\{1,...,L\}$ can be calculated via,
\begin{equation}
R_{i}=[\lvert y_1^{\rm{cal}}-\hat{y}_1^{\rm{cal}}\rvert,...,\lvert y_D^{\rm{cal}}-\hat{y}_D^{\rm{cal}}\rvert]^{\top}.
\label{Eq.5}
\end{equation}

We denote the nonconformity scores of $L$ samples as $R\in \mathbb{R}^{L\times D}$. And we use $\boldsymbol{\hat{\epsilon}}=[\hat{\epsilon}_1,...,\hat{\epsilon}_D] \in \mathbb{R}^{D}$ to represent the empirical nonconformity score. For the $i,i\in \{1,...,D\}$-th dimension, all $L$ samples are sorted and the $l,l=\left\lceil {(1-\alpha)(L+1)} \right\rceil$-th smallest volume is chosen as the $i$-th element of the empirical nonconformity score $\hat{\epsilon}_i$\cite{CF-RNN}. 

For a new example in the testing set, we can obtain the prediction of our proposed model with the following formulation,
\begin{equation}
\boldsymbol{\hat{y}}=f_{\boldsymbol{\theta} ^ {*}}(\boldsymbol{X}^{\rm{tra,ts}},\boldsymbol{X}^{\rm{txt,ts}},\boldsymbol{X}^{\rm{img}}),
\label{Eq.6}
\end{equation}
where $\hat{\boldsymbol{y}}=[\hat{y}_1,...,\hat{y}_D]$ and $\hat{y}_i$ denotes the $i$-th element of $\boldsymbol{\hat{y}}$. 

Then, the prediction interval can be obtained by the following equations, 
\begin{align}
\hat{y}_i^{\rm{L}}=\hat{y}_{i}-\hat{\epsilon}_i, \label{Eq.7}\\
\hat{y}_i^{\rm{U}}=\hat{y}_{i}+\hat{\epsilon}_i, \label{Eq.8}
\end{align}
where $\hat{\boldsymbol{y}}^{\rm{L}}=[\hat{y}_1^{\rm{L}},...,\hat{y}_D^{\rm{L}}]$,${\hat{\boldsymbol{y}}}^{\rm{U}}=[\hat{y}_1^{\rm{U}},...,\hat{y}_D^{\rm{U}}]$ denote the lower bound and upper bound of the prediction interval. $\hat{y}_i^{\rm{L}}$ and $\hat{y}_i^{\rm{U}}$ denote the $i$-th element of $\hat{\boldsymbol{y}}^{\rm{L}}$ and $\hat{\boldsymbol{y}}^{\rm{U}}$, respectively.

Finally, when we fix a significance level $\alpha\in (0,1)$, the probability that the true volume falls outside the corresponding prediction interval is at most $\alpha$ \cite{jensen2022ensemble,CQR},
\begin{equation}
\mathbb{P}(y_i \in [\hat{y}_i^{\rm{L}},\hat{y}_i^{\rm{U}}])\geq 1-\alpha.
\label{Eq.9}
\end{equation}

\subsection{Model Description of MetaSTNet}
We propose a meta-learning based spatiotemporal neural network entitled MetaSTNet. The training procedure of our proposed model is shown in Algorithm 1. Lines 1-15 demonstrate the meta-training phase, which aims to train the model and obtain the favorable initialized parameters. While lines 16-26 show the procedures of point prediction and constructing prediction intervals. 

\begin{figure}[!t]
	\centering          
	\includegraphics[width=\columnwidth]{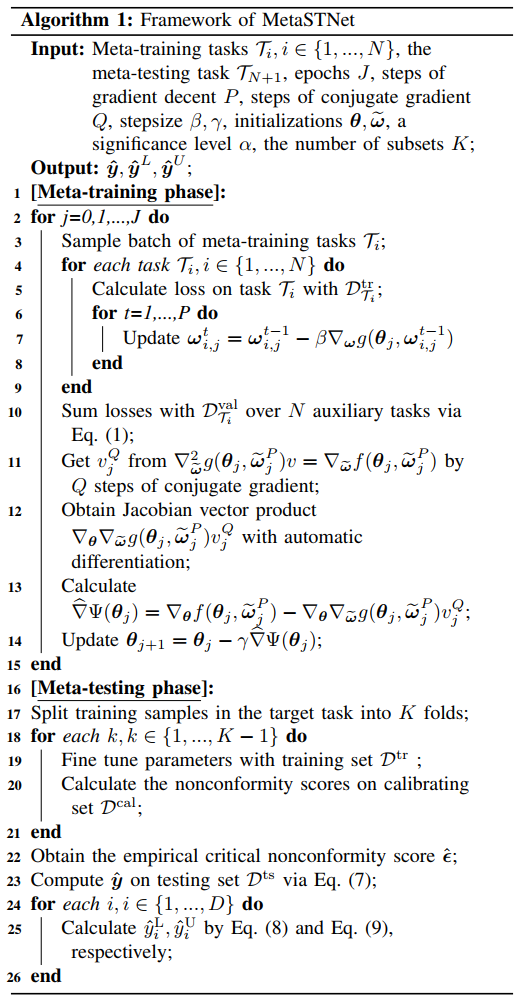}
	\label{alg1}
\end{figure}
  
\begin{figure*}[!t]
\centering          
\includegraphics[width=7in]{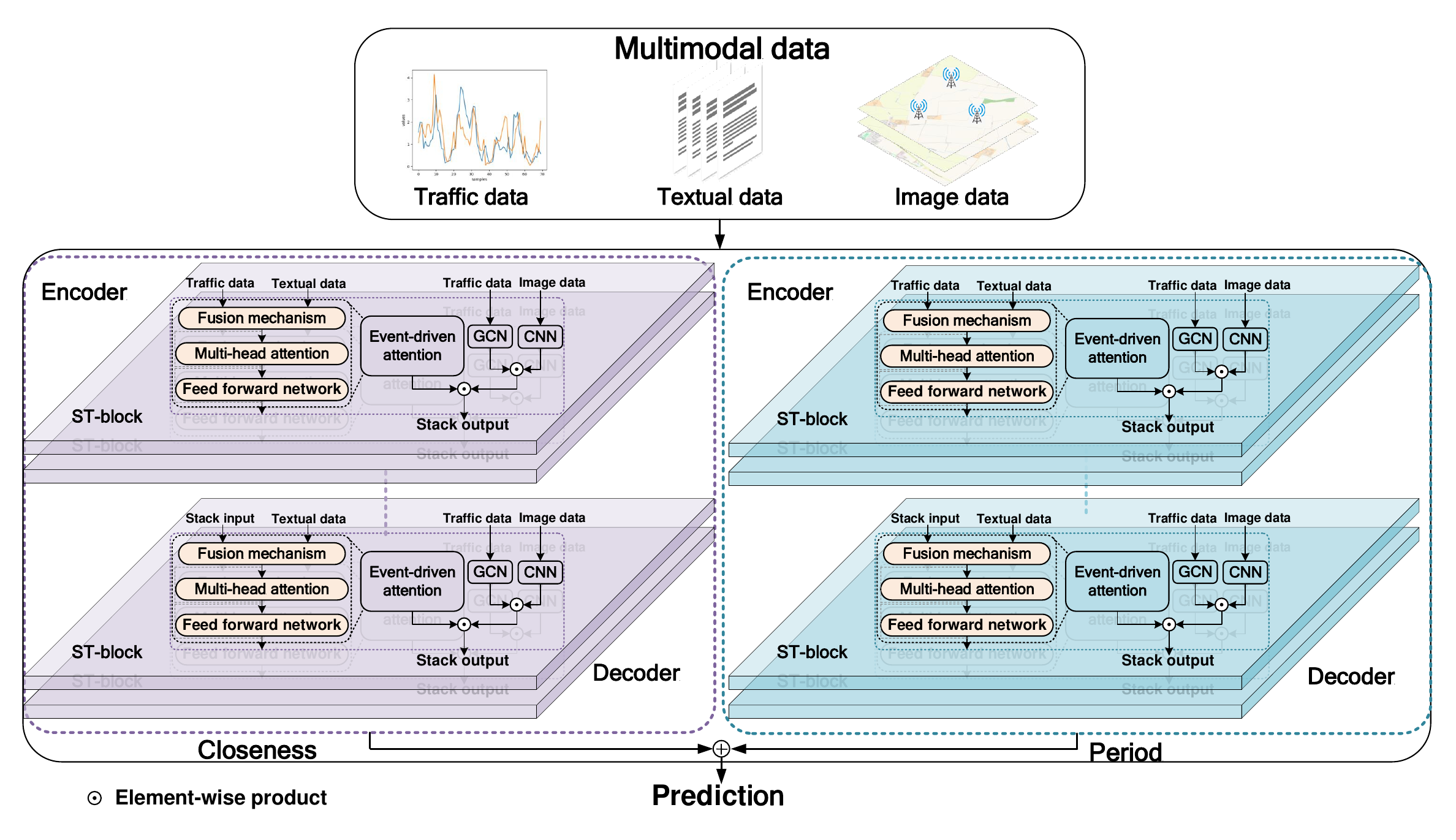}
\caption{The structure of the proposed MetaSTNet.}
\label{MetaSTNet}
\end{figure*}

Fig. \ref{MetaSTNet} illustrates the structure of MetaSTNet. Specifically, we use data in three modalities, including traffic, textual, and image data. MetaSTNet consists of two parallel encoder-decoder structures, which aim to capture the short-term and periodic features, respectively. At the same time, each encoder-decoder structure is composed of a stack of ST-blocks for mining the spatiotemporal correlations of multimodal data. Furthermore, in each ST-block, we design event-driven attention to further extract the long-term temporal dependencies of traffic sequences and textual data. Next, the building blocks of MetaSTNet will be explained in detail. For notational simplicity, we omit the subscript $\mathcal{T}_i$ when denoting the data and parameters of task $\mathcal{T}_i$.

\subsubsection{Event-driven attention} 
Related studies\cite{Zhou2022,Lim2019,Liu2022a} show that the multi-head attention mechanism helps to extract the long-term time-domain dependencies of time series, as it can eliminate error accumulation and use multiple heads to capture features of different aspects of time series for a better representation\cite{NEURIPS2020_c6b8c8d7,xueye_2022}. Based on this, we design event-driven attention to capture long-term time-domain dependencies from traffic data and textual information (e.g., social activities and news events).

The input of event-driven attention includes traffic and textual data, which can be represented as $\boldsymbol{H}_{\rm{tra}}$ and $\boldsymbol{H}_{\rm{txt}}$. The traffic data $\boldsymbol{H}_{\rm{tra}}$ in the first layer of ST-block is obtained by Embedding operation on $\boldsymbol{X}^{\rm{tra}}$, while $\boldsymbol{H}_{\rm{txt}}$ is obtained by Embedding operation on $\boldsymbol{X}^{\rm{txt}}$.

To uncover the complementary information of these two modalities, we integrate the long-term time-domain features with the fusion mechanism, which can be represented as,
\begin{align}		
\boldsymbol{Z} &= Sigmoid(\boldsymbol{H}_{\rm{tra}}+\boldsymbol{H}_{\rm{txt}}),\\
\boldsymbol{H}^{'} &=\boldsymbol{Z} \boldsymbol{H}_{\rm{tra}}+(1-\boldsymbol{Z}) \boldsymbol{H}_{\rm{txt}}.
\end{align}

Then, we conduct multi-head attention mechanism to capture the long-term time-domain features. Specifically, the query, key, and value matrices are denoted as $\boldsymbol{Q},\boldsymbol{K},\boldsymbol{V}$, respectively. They can be obtained via matrix multiplication between $\boldsymbol{H}^{'}$ and learnable weights, namely $\boldsymbol{W}^{Q},\boldsymbol{W}^{K},\boldsymbol{W}^{V}$,
\begin{equation}\boldsymbol{Q}=\boldsymbol{H}^{'} \boldsymbol{W}^{Q}, 
\boldsymbol{K}=\boldsymbol{H}^{'} \boldsymbol{W}^{K},
\boldsymbol{V}=\boldsymbol{H}^{'} \boldsymbol{W}^{V}.\label{eq}\end{equation}

The attention is based on the scaled dot product\cite{attention} and it can be computed as, 
\begin{equation}
Attention(\boldsymbol{Q},\boldsymbol{K},\boldsymbol{V})=Softmax(\frac{\boldsymbol{Q}\boldsymbol{K}^{T}}{\sqrt{d_{a}}}\boldsymbol{V}). \label{att}
\end{equation}

The abovementioned attention process is performed multiple times in parallel\cite{9930825}. Suppose there are $n$ heads and $\boldsymbol{h}_{i},i\in \{1,...,n\}$ denotes the $i$-th number of attention heads. And it can be obtained with Eq. (\ref{Eq.14}),
\begin{equation}
\boldsymbol{h}_{i}=Attention(\boldsymbol{Q} \boldsymbol{W}_i^{\rm{q}},\boldsymbol{K} \boldsymbol{W}_i^{\rm{k}},\boldsymbol{V} \boldsymbol{W}_i^{\rm{v}}),
\label{Eq.14}
\end{equation}
where $\boldsymbol{W}_i^{\rm{q}},\boldsymbol{W}_i^{\rm{k}},\boldsymbol{W}_i^{\rm{v}}$ denote the learnable weights of the query, key, and value of the $i$-th attention head, respectively. Then, the output of multi-head attention can be achieved via Eq. (\ref{Eq.15}),
\begin{equation}
Multihead(\boldsymbol{Q},\boldsymbol{K},\boldsymbol{V})=Concat(\boldsymbol{h}_1,\boldsymbol{h}_2,...,\boldsymbol{h}_n)\boldsymbol{W}_{0},
\label{Eq.15}
\end{equation}
where $\boldsymbol{W}_0$ is the learnable weight.

The last component of event-driven attention is FFN. It is composed of the linear transformations with Relu activation function, i.e.,
\begin{equation}
FFN(\boldsymbol{x})=Relu(0,\boldsymbol{x}\boldsymbol{W}_1+\boldsymbol{b}_1)\boldsymbol{W}_2 + \boldsymbol{b}_2,
\end{equation}
where $\boldsymbol{W}_1,\boldsymbol{W}_2,\boldsymbol{b}_1,\boldsymbol{b}_2$ are learnable parameters and $\boldsymbol{x}$ represents the output of multihead attention after the Add \& Normalization layer\cite{attention}. Finally, we denote the output of the FFN as $\boldsymbol{H}_{\rm{temporal}}\in \mathbb{R}^{T\times d_t}$.

\subsubsection{ST-block} Cellular traffic data has complex temporal and spatial correlations\cite{b6,TWACNet,GCLSTM} and is influenced by multiple external factors. For instance, social activities and the spatial distribution of urban buildings are directly related to the generation of cellular traffic\cite{STCNet}. For this reason, we employ a stack of ST-blocks to capture the spatiotemporal correlations of multimodal data (including traffic, textual, and image data). In each ST-block, event-driven attention is adopted to capture the long-term temporal dependencies. At the same time, GCN and CNN capture the spatial correlations of traffic and image data, respectively.

Let $f_{\rm{gcn}}(\cdot),f_{\rm{cnn}}(\cdot)$ denote GCN and CNN, respectively. $\boldsymbol{H}_{\rm{gcn}},\boldsymbol{H}_{\rm{cnn}}$ denote the features extracted from GCN and CNN, respectively. $G \in \mathbb{R}^{D \times D}$ represents the coefficient matrix of spatial correlations based on geographic location. Then, the spatial features $\boldsymbol{H}_{\rm{gcn}},\boldsymbol{H}_{\rm{cnn}}$ can be obtained via,
\begin{align}
\boldsymbol{H}_{\rm{gcn}} &= f_{\rm{gcn}}(\boldsymbol{X}^{\rm{tra}},G),\\
\boldsymbol{H}_{\rm{cnn}} &= f_{\rm{cnn}}(\boldsymbol{X}^{\rm{img}}).
\end{align}

Besides, we combine the temporal and spatial dependencies by the element-wise product, which is shown as follows,
\begin{align} 
&\boldsymbol{H}_{\rm{spatial}} = 
\boldsymbol{H}_{\rm{gcn}} \odot \boldsymbol{H}_{\rm{cnn}},\label{Eq.19}\\
&\boldsymbol{H}_{\rm{st}}=\boldsymbol{H}_{\rm{temporal}}\odot \boldsymbol{H}_{\rm{spatial}},\label{Eq.20}
\end{align}
where $\boldsymbol{H}_{\rm{st}}$ denotes the output of a ST-block (namely stack output).

\subsubsection{Encoder-decoder} 
Previous studies \cite{DeseNet,ST-Tran,Liu2019} explored that cellular network traffic has multi-scale time-domain characteristics. For example, the recent history captures immediate changes in traffic data, while the periodicity extracts the global trends. Inspired by this, we design two paralleled encoder-decoder structures to model the closeness and period of traffic sequences, respectively. Let $\boldsymbol{H}^{\rm{c}}_{\rm{st}},\boldsymbol{H}^{\rm{p}}_{\rm{st}}$ denote the output of the short-term and periodic components, respectively. $\boldsymbol{w}_1,\boldsymbol{w}_2,\boldsymbol{b}$ are learnable parameters. Then we can obtain the predicted volumes $\boldsymbol{\hat{y}}$ by the following equation,
\begin{equation}
\boldsymbol{\hat{y}} = \boldsymbol{w}_1 \boldsymbol{H}^{\rm{c}}_{\rm{st}} + \boldsymbol{w}_2 \boldsymbol{H}^{\rm{p}}_{\rm{st}} + \boldsymbol{b}.
\end{equation}

\section{Experimental Setups}
\subsection{Dataset Description}
To evaluate the performance of our proposed model, we leverage the simulation data generated from a simulator and three real-world datasets (i.e., Milano, Trento, and LTE traffic data) to carry out the comparison experiments of cellular traffic prediction methods.
\subsubsection{Simulator} We develop a traffic simulator that can generate network traffic in multicell/multiuser environments. The traffic simulator consists of a set of base stations and each base station comprises three sectors with many users per sector. For each user, we generate traffic data with seasonal and random terms. In addition, we design an intercell handover procedure to simulate user mobility during working time. The mobile users randomly select the direction of movement and move towards the simulation boundary region. They change the direction when the simulated boundary region is reached. Finally, the traffic volumes for each cell can be obtained by collecting all users' traffic data in that cell.

\subsubsection{Milano} 
The Milano dataset collects heterogeneous multimodal data, including telco, textual, and image data from Milan city, ranging from 1 November 2013 to 1 January 2014. The whole spatial space is composed of 10000 grids with an approximate size of 235$\times$235 square meters in each grid\cite{Barlacchi2015}. The telco data\cite{milan_data} records three types of call detail records, i.e., short message service, call service, and Internet service. This paper focuses on network traffic data at a time granularity of one hour. The textual data is made up of social pulse and daily news datasets. The social pulse dataset\cite{Social_Pulse_milan} contains anonymized usernames, geographical coordinates, and geolocalized tweets, while the daily news dataset\cite{milanotoday} contains the information such as titles, topics, and types of news. The image data records the traffic road network and building blocks of the city, which can be collected from OpenStreetMap\footnotemark[1].
    
\subsubsection{Trento} 
The Trento dataset collects heterogeneous data with three modalities (i.e., traffic, textual, and image data) in the province of Trentino. The telco data can be obtained via\cite{trentino_data}. The textual data, including social pulse and website news data, are also publicly available, which can be obtained through\cite{Social_Pulse_trentino, TrentoToday}, respectively. Besides, the image data is also collected from OpenStreetMap\footnotemark[1] when given the spatial location (longitude and latitude).

\subsubsection{LTE traffic}		
The LTE traffic dataset is collected from a private operator. It records the downlink traffic data for two weeks with a temporal interval of one hour. We use downlink traffic, downlink throughput, and timestamps as auxiliary data. In terms of image data, we can also get an open-source map from OpenStreetMap\footnote{www.openstreet.com} with latitude and longitude information. 

\subsection{Compared Methods}
\begin{enumerate}
\item \emph{GRU:} It is a sequential model using GRU cells, which is of a simple structure and has been widely used in time series prediction.	
\item \emph{LSTM:} A better variant of recurrent neural networks can tackle this problem of long-range time-domain dependencies that RNNs cannot handle.	
\item \emph{GCN\cite{welling2016semi}:} This architecture is utilized to capture the spatial dependencies of traffic data.
\item \emph{Transformer\cite{attention}:} It uses the attention mechanism to model long-term and short-term dependencies of cellular traffic.		
\item \emph{DeseNet\cite{DeseNet}:} This architecture utilizes paralleled CNN to adaptively capture spatiotemporal correlations of cellular traffic among neighboring cells. 		
\item \emph{STCNet\cite{STCNet}:} It uses ConvLSTM and CNN to model spatiotemporal dependencies and exogenous factors, respectively.		
\item \emph{DeepAuto\cite{DeepAuto}:} It uses three paralleled LSTM to capture the trend, period, and closeness traffic patterns. Besides, auxiliary factors have been utilized to improve prediction performance.		
\item \emph{ST-Tran\cite{ST-Tran}:} It utilizes transformer and GCN to characterize the spatiotemporal relationships of cellular traffic.		
\item \emph{dmTP\cite{Zhang2021-dmTP}:} It utilizes LSTM to capture long-term time-domain dependencies with a model-based meta-learning framework.
\end{enumerate}

Besides, we design three variants of MetaSTNet, which are shown as follows.
\begin{enumerate}
\item \emph{MetaSTNet/oExt\_oMeta:} MetaSTNet without external information and meta-learning framework.	\item \emph{MetaSTNet/oExt:} MetaSTNet without external information.
\item \emph{MetaSTNet/oMeta:} MetaSTNet without meta-learning framework.
\end{enumerate}

\subsection{Evaluation Metrics}
We use MAE and RMSE to evaluate the point prediction performance of different models. Let $M$ denote the number of values, MAE and RMSE can be achieved via,
\begin{align}
\text{MAE} &= \frac{1}{M}\sum_{i=1}^{M}|\boldsymbol{y}_i-\boldsymbol{\hat{y}}_i|,\\
\text{RMSE} &=  \sqrt{\frac{1}{M}\sum_{i=1}^{M}(\boldsymbol{y}_i-\boldsymbol{\hat{y}}_i)^{2}}, 
\end{align}
where $\boldsymbol{y}_i$ and $\boldsymbol{\hat{y}}_i$ are the $i$-th elements of the target and predicted volumes, respectively.

Besides, we use coverage rate (CR) and average width length (WL)\cite{Gupta2022} to evaluate the performance of interval prediction, which can be represented as,
\begin{align}
\text{CR} &= \frac{1}{MD} \sum_{i=1}^{M} \sum_{j=1}^{D} \mathbbm{1}\left\{\hat{y}_{i,j}^{\rm{L}} \leq y_{i,j} \leq \hat{y}_{i,j}^{\rm{U}}\right\},  \\
\text{WL} &=\frac{1}{MD} \sum_{i=1}^{M} \sum_{j=1}^{D}|\hat{y}_{i,j}^{\rm{U}}-\hat{y}_{i,j}^{\rm{L}}|,
\end{align}
where $y_{i,j}$ is the true volume, while $\hat{y}^{\rm{L}}_{i,j}$ and $\hat{y}^{\rm{U}}_{i,j}$ denote the lower bound and upper bound of a prediction interval, respectively.

\subsection{Experimental Details}
\begin{enumerate}
\item \emph{Preprocessing:} The Min-Max normalization is utilized to scale numerical data to [0, 1]. Furthermore, we transform metadata such as holidays, the day of the week, and the day of the hour via one-hot encoding.
\item \emph{Hyperparameter Setting:} In our study, we set $N$ to 4, i.e., one target task and four auxiliary tasks. Moreover, a larger number of neighboring cells and auxiliary tasks can also be considered if the amount of data is sufficient. For each auxiliary task, we select simulation data as the training set (support set) and real-world traffic data as the validating set (query set). For the target task, we reckon cellular traffic from the last seven days as the testing data and all traffic before as training and calibrating data. In addition, we set horizon $h = \{1, 24\}$ to forecast traffic for the next hour and the next day, respectively. The other hyper-parameter settings are listed as follows: learning rate is 0.001, the dropout rate is 0.05, the dimensions of hidden layers are 128, the number of the head is 8, while the number of ST-blocks is 2, the timestep for close and period data are both 3 for one-hour ahead prediction and 6 for one-day ahead prediction. The number of epochs for meta-training is 2000. 
\item \emph{Experimental environment:} We conduct all experiments on a Linux server with four 12GB GPUs with NVIDIA TITAN X (Pascal). Besides, we use the deep learning framework of Pytorch 1.6.0 with the programming language of python 3.7.
\end{enumerate}

\section{Experimental Results}
\subsection{Point Prediction}
\begin{table*}[t]
\centering
\caption{prediction performance comparisons in terms of MAE and RMSE on three real-world datasets}
\label{tab:point_prediction}
\renewcommand\arraystretch{1.25}
\resizebox{\linewidth}{!}{
\begin{tabular}{c|cccc|cccc|cccc|c}
\toprule
Datasets                 & \multicolumn{4}{c|}{Milano}                                                                                     & \multicolumn{4}{c|}{Trento}                                                                                      & \multicolumn{4}{c|}{LTE traffic}                                                                                   & \multirow{3}{*}{Average\_rank} \\ \cline{1-13}
\multirow{2}{*}{Methods} & \multicolumn{2}{c|}{1-hour}                                      & \multicolumn{2}{c|}{1-day}                  & \multicolumn{2}{c|}{1-hour}                                        & \multicolumn{2}{c|}{1-day}                    & \multicolumn{2}{c|}{1-hour}                                      & \multicolumn{2}{c|}{1-day}                  &                                \\ \cline{2-13}
                     & MAE                  & \multicolumn{1}{c|}{RMSE}                 & MAE                  & RMSE                 & MAE                   & \multicolumn{1}{c|}{RMSE}                  & MAE                   & RMSE                  & MAE                  & \multicolumn{1}{c|}{RMSE}                 & MAE                  & RMSE                 &                                \\ \midrule
GRU                      & 2.793                & \multicolumn{1}{c|}{3.859}                & 3.415                & 4.641                & 13.536                & \multicolumn{1}{c|}{22.302}                & 16.356                & 25.468                & 0.427                & \multicolumn{1}{c|}{0.582}                & 0.542                & 0.704                & 10.17                          \\
LSTM                     & 2.788                & \multicolumn{1}{c|}{3.886}                & 3.383                & 4.667                & 14.017                & \multicolumn{1}{c|}{23.674}                & 16.217                & 25.270                & 0.469                & \multicolumn{1}{c|}{0.630}                & 0.535                & 0.695                & 11.08                          \\
Transformer              & 2.662                & \multicolumn{1}{c|}{3.796}                & 3.508                & 4.669                & 13.869                & \multicolumn{1}{c|}{23.027}                & 16.688                & 25.613                & 0.420                & \multicolumn{1}{c|}{0.574}                & 0.511                & 0.677                & 9.25                           \\
GCN                      & 2.780                & \multicolumn{1}{c|}{3.910}                & 3.494                & 4.721                & 14.246                & \multicolumn{1}{c|}{23.465}                & 16.446                & 26.465                & 0.450                & \multicolumn{1}{c|}{0.614}                & 0.589                & 0.787                & 11.83                          \\
DeseNet                  & 2.711                & \multicolumn{1}{c|}{3.792}                & 3.287                & 4.466                & 13.866                & \multicolumn{1}{c|}{23.082}                & 16.768                & 26.589                & 0.458                & \multicolumn{1}{c|}{0.620}                & 0.523                & 0.714                & 10.67                          \\
STCNet                   & 2.690                & \multicolumn{1}{c|}{3.601}                & 3.130                & 4.192                & 13.468                & \multicolumn{1}{c|}{22.272}                & 15.908                & 24.925                & 0.460                & \multicolumn{1}{c|}{0.639}                & 0.511                & 0.699                & 7.75                           \\
DeepAuto                 & 2.693                & \multicolumn{1}{c|}{3.677}                & 3.149                & 4.164                & 13.406                & \multicolumn{1}{c|}{22.171}                & 15.521                & 25.188                & 0.413                & \multicolumn{1}{c|}{0.548}                & 0.514                & 0.659                & 6.33                           \\
ST-Tran                 & 2.634                & \multicolumn{1}{c|}{3.687}                & 3.068                & 3.917                & 12.859                & \multicolumn{1}{c|}{21.612}                & 15.683                & 25.255                & 0.408                & \multicolumn{1}{c|}{0.575}                & 0.516                & 0.668                & 5.92                           \\
dmTP                     & 2.687                & \multicolumn{1}{c|}{3.648}                & 3.259                & 4.226                & 13.145                & \multicolumn{1}{c|}{21.765}                & 15.645                & 25.047                & 0.452                & \multicolumn{1}{c|}{0.606}                & 0.510                & 0.676                & 6.75                           \\ \hline
MetaSTNet/oExt\_oMeta   & 2.684                & \multicolumn{1}{c|}{3.658}                & 3.048                & 4.179                & 12.715                & \multicolumn{1}{c|}{21.940}                & 15.138                & 24.760                & 0.403                & \multicolumn{1}{c|}{0.561}                & 0.499                & 0.667                & 4.50                           \\
MetaSTNet/oMeta         & 2.515                & \multicolumn{1}{c|}{3.619}                & 3.062                & 3.989                & 12.616                & \multicolumn{1}{c|}{21.474}                & 14.984                & 24.164                & 0.391                & \multicolumn{1}{c|}{0.534}                & 0.495                & 0.651                & 2.75                           \\
MetaSTNet/oExt          & 2.498                & \multicolumn{1}{c|}{3.658}                & 2.881                & 3.911                & 12.752                & \multicolumn{1}{c|}{20.312}                & 15.030                & 22.613                & 0.401                & \multicolumn{1}{c|}{0.528}                & 0.501                & 0.653                & 2.83                           \\
MetaSTNet               & {\ul \textbf{2.481}} & \multicolumn{1}{c|}{{\ul \textbf{3.526}}} & {\ul \textbf{2.778}} & {\ul \textbf{3.807}} & {\ul \textbf{12.602}} & \multicolumn{1}{c|}{{\ul \textbf{19.063}}} & {\ul \textbf{14.778}} & {\ul \textbf{21.210}} & {\ul \textbf{0.387}} & \multicolumn{1}{c|}{{\ul \textbf{0.521}}} & {\ul \textbf{0.490}} & {\ul \textbf{0.645}} & 1.00                           \\ \bottomrule
\end{tabular}}
\end{table*}

We evaluate the performance of thirteen models for one-hour and one-day ahead prediction on three real-world datasets. Our proposed model can achieve significantly superior prediction performance with only a small amount of real-world data (e.g., two weeks of traffic data). The baseline models in our study, however, perform exceptionally poorly. Therefore, we conduct experiments with one-month real-world traffic data to fully explore the potential of those baselines and compare their prediction performance more fairly, at the expense of reducing the performance gains of MetaSTNet to some degree. Specifically, we use one-week real-world traffic samples as the testing set and the previous 23 days of traffic data as the training set. The best results are underlined in bold. Moreover, we use the average rank to measure the prediction performance of different models. 

As is shown in Table \ref{tab:point_prediction}, GRU and LSTM can capture the long-term time-domain dependencies of time series with a gated structure. However, these methods cannot capture the spatial dependencies of traffic data, resulting in poor performance. Transformer can capture temporal dependencies for very long periods with multi-head attention mechanism but cannot capture spatial dependencies. In contrast, GCN is adopted to capture the local spatial dependencies, but it performs poorly as it cannot model the time-domain characteristics of traffic data. DeseNet uses three residual networks to model hourly, daily, and weekly traffic patterns and capture the spatiotemporal correlations of fine-grained traffic data. 

STCNet utilizes LSTM and CNN to capture the long-term temporal dependencies and local spatial correlations of traffic sequences, respectively. Besides, it can effectively fuse external information and performs significantly better than the above five models. In order to take the short-term, periodic, and trend patterns into consideration, DeepAuto utilizes three paralleled LSTM to model the closeness, period, and trend traffic data, respectively. Besides, the auxiliary information is further considered in DeepAuto. The results show that it performs better than STCNet. Moreover, ST-Tran is a spatiotemporal structure that can effectively capture the spatiotemporal dependencies of traffic data. The prediction performance is relatively optimal among nine comparative baselines. While dmTP is a deep learning model based on a meta-learning framework for network traffic prediction. It adopts LSTM as a base model. However, it cannot capture spatial dependencies, making the performance of dmTP not as good as our proposed model.

The prediction performance of MetaSTNet is relatively optimal compared to the above state-of-the-art approaches. There are several possible explanations for this. To begin with, we can obtain the meta-knowledge from simulation data and transfer the learned meta-knowledge to the real environment. With a bi-level optimization procedure, the model has the ability to adapt quickly to the target task and achieve superior prediction performance. Besides, we design event-driven attention to capture the long-term temporal correlations and model the spatial relationships with GCN and CNN. Also, we introduce ST-block to fuse spatiotemporal dependencies of multimodal data. Finally, we design two paralleled encoder-decoder structures to model short-term (hourly) and periodic (daily) multimodal data, respectively. In this way, the prediction accuracy can be further improved.


\subsection{Ablation Study}
We show the point prediction results of the ablation experiment on three real-world datasets. As can be observed in Table \ref{tab:point_prediction} that all three variants of MetaSTNet (i.e., MetaSTNet/oExt\_oMeta, MetaSTNet/oMeta, MetaSTNet/oExt) perform worse than MetaSTNet, indicating that removing the meta-learning framework or external information will reduce the prediction accuracy of MetaSTNet to a certain extent. Among the three variants, MetaSTNet/oExt\_oMeta performs worst for two reasons. On the one hand, it uses unimodal data and the feature representation is still insufficient, which affects the point prediction performance. On the other hand, it is challenging to obtain ideal prediction performance without a meta-learning framework, as the model can only be trained with a small amount of real-world traffic data. Therefore, we can conclude from the ablation study that both auxiliary information and meta-learning framework play a significant role in obtaining superior point prediction performance.

\subsection{Experiment with Simulation Data}

\begin{table*}[t]
\centering
\caption{prediction performance comparisons in different proportions of synthetic and real-world datasets}
\label{tab:synthesis}
\renewcommand{\arraystretch}{1.5}
\begin{tabular}{c|cccc}
\toprule
\multirow{2}{*}{Ratio of simulation and real-world datasets} & \multicolumn{2}{c|}{1-hour} & \multicolumn{2}{c}{1-day}  \\ \cline{2-5}  & MAE   & \multicolumn{1}{c|}{RMSE}   & MAE  & RMSE   \\ \midrule
Real-world data only & 2.530   & \multicolumn{1}{c|}{3.553} & 2.838  & 3.794   \\
1:1  &  2.547   &   \multicolumn{1}{c|}{3.573}  & 2.864 & 4.056 \\
2:1  &  2.495   &   \multicolumn{1}{c|}{3.545}  & 2.804 & 3.829 \\
3:1  &  2.481   &   \multicolumn{1}{c|}{3.526}  & 2.778 & 3.807 \\
4:1  & {\ul \textbf{2.470}}  & \multicolumn{1}{c|}{{\ul \textbf{3.351}}}  & {\ul \textbf{2.751}}  & {\ul \textbf{3.769}} \\
8:1   & 2.475   &\multicolumn{1}{c|}{3.531}  & 2.752 & 3.777    \\ \bottomrule
\end{tabular}
\end{table*}

In this section, we explore the impact of different amounts of simulation data on the prediction accuracy of MetaSTNet during the meta-training phase. For a better illustration, we conduct experiments on the Milano dataset and use one-week real-world traffic data as the validating set in the meta-training stage. Specifically, "Real-world data only" means that we use one-week real-world training samples as the training set in the meta-training phase, while 1:1, 2:1, and 3:1 mean that we train MetaSTNet with one week, two weeks, three weeks simulation data as the training set in the meta-training stage, respectively. Besides, 4:1 and 8:1 can roughly consider that we train MetaSTNet using one month and two months of simulation data as the training set in the meta-training stage. 

The experimental result is shown in Table \ref{tab:synthesis}. It can be observed from Table \ref{tab:synthesis} that, when the number of real-world training samples is very small, simulation data is shown to be beneficial and can improve the prediction accuracy to some extent. This is because a small number of real-world training samples can only provide a few insights. In contrast, synthetic data can give more useful information, which will help to capture the complex spatiotemporal dependencies of network traffic. Besides, the prediction accuracy improves as the ratio of synthetic data gradually increases, and the prediction performance is relatively optimal when the ratio reaches 4:1. As the ratio continues to increase, the prediction accuracy of MetaSTNet remains stable. Overall, the experiments demonstrate that it is feasible to enhance the prediction performance of cellular network traffic to leverage synthetic data when only a very small number of training samples are available.

\subsection{Interval Prediction}
In this section, we compare the performance of interval prediction based on inductive and cross conformal prediction within different confidence levels. We refer to the methods of using inductive conformal prediction and cross conformal prediction as MetaSTNet\_ICP and MetaSTNet\_CCP for short, respectively. Coverage rate and average width length\cite{Gupta2022} can be used to evaluate the performance of interval prediction with a predetermined confidence level. 

\begin{figure*}[!t]	
\centering{\includegraphics[width=7.2in]{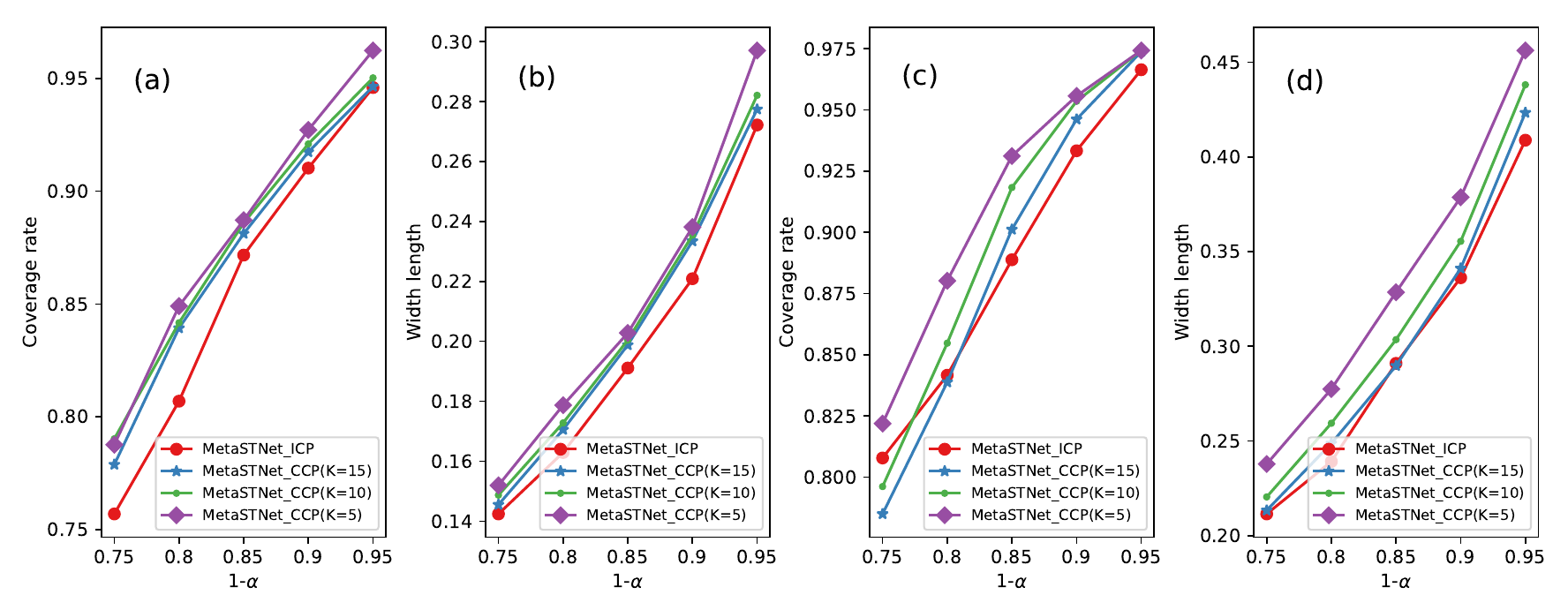}}
\caption{The interval prediction results on the Milano dataset. Among four subplots, (a) and (b) display 1-hour ahead interval prediction results with metrics of coverage rate and average width length, respectively. In contrast, (c) and (d) show the 1-day ahead interval prediction results with metrics of coverage rate and average width length, respectively.}
\label{milan}
\end{figure*}

\begin{figure*}[!t]	
\centering{\includegraphics[width=7.2in]{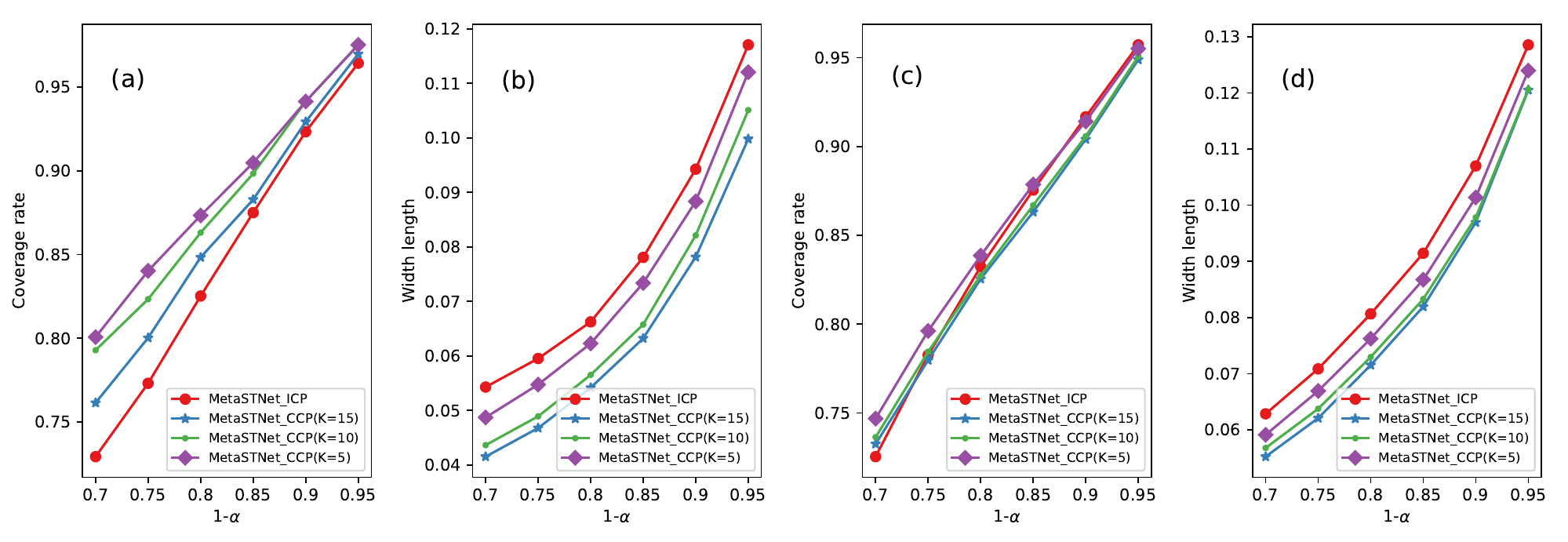}}
\caption{The interval prediction results on the Trento dataset. Among four subplots, (a) and (b) display 1-hour ahead interval prediction results with metrics of coverage rate and average width length, respectively. In contrast, (c) and (d) show the 1-day ahead interval prediction results with metrics of coverage rate and average width length, respectively.}
\label{trento}
\end{figure*}

Fig. \ref{milan} evaluates the performance of short-term (1-hour ahead) and long-term (1-day ahead) interval forecasting on the Milano dataset. These four subplots explore the effect of $K$ on interval prediction performance under different coverage rates. When $K=2$, MetaSTNet\_CCP is equivalent to MetaSTNet\_ICP, which can be reckoned as a special case of MetaSTNet\_CCP. We can see from Fig. \ref{milan}(a) that MetaSTNet\_CCP with $K=5$ can obtain the optimal coverage rate compared with MetaSTNet\_ICP, MetaSTNet\_CCP ($K=10$), and MetaSTNet\_CCP ($K=15$). In MetaSTNet\_ICP, only half of the training data can be used to calibrate nonconformity scores. Thus, a limited amount of calibrating data may lead to a high variance of confidence. In order to use more training data for calibration, we choose $K$ with a higher number. When $K$ is equal to 5, the coverage rate and average width length of MetaSTNet\_CCP is higher than that of MetaSTNet\_ICP, indicating that a higher $K$ helps improve coverage rate. However, when $K$ reaches 10 or 15, the interval prediction performance is not as good as MetaSTNet\_CCP ($K=5$), which shows that a too large $K$ will deteriorate interval prediction performance. Moreover, we can see from Fig. \ref{milan}(c) and Fig. \ref{milan}(d) that the coverage rate and the width length of MetaSTNet\_CCP ($K=5$) are higher than that of MetaSTNet\_ICP, indicating that MetaSTNet\_CCP ($K=5$) can contain more true volumes in the long-term interval prediction.

Furthermore, Fig. \ref{trento} evaluates the performance of short-term (1-hour ahead) and long-term (1-day ahead) interval prediction on the Trento dataset. It can be seen from Fig. \ref{trento}(a) and Fig. \ref{trento}(b) that the coverage rate of MetaSTNet\_CCP ($K=5$) is significantly high than that of MetaSTNet\_ICP and the average width length of MetaSTNet\_CCP ($K=5$) is slightly lower than MetaSTNet\_ICP, indicating that MetaSTNet\_CCP performs better than MetaSTNet\_ICP when $K$ is equal to 5. In terms of 1-day ahead interval prediction, it can also be observed from Fig. \ref{trento}(c) and Fig. \ref{trento}(d) that MetaSTNet\_CCP ($K=5$) performs slightly better than MetaSTNet\_ICP. 

\subsection{Visualization Analysis}

\begin{figure}[!t]		
\centering{\includegraphics[width=\columnwidth]{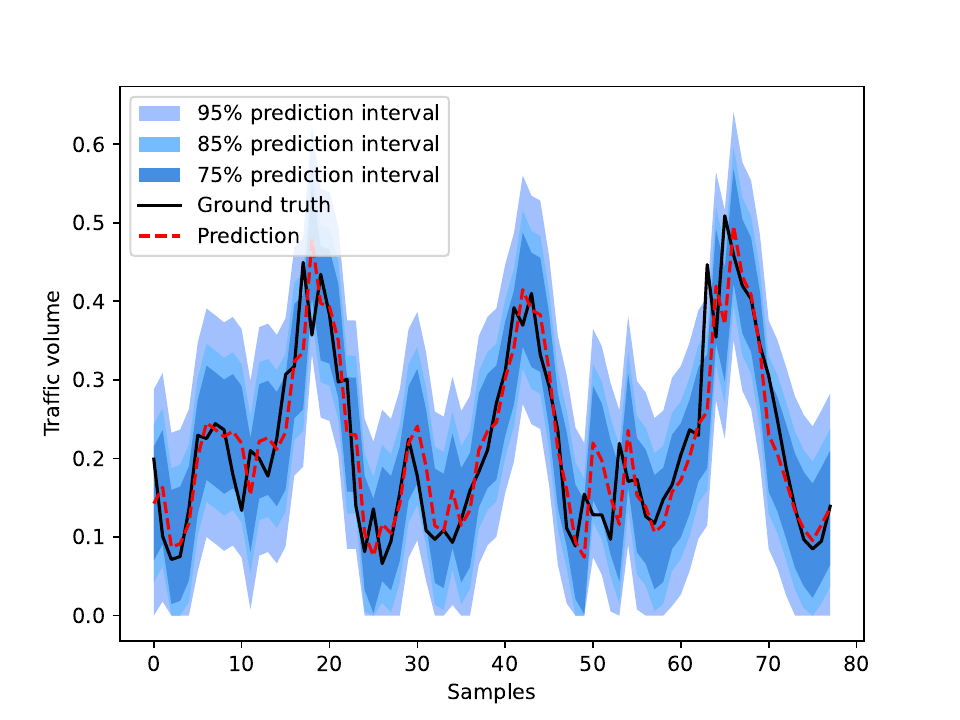}}
\caption{The experimental results of 1-hour ahead interval prediction on the Milano dataset. The black line denotes ground truths and the red line represents predictions. The gradually deepening blue colors represent the prediction intervals under the confidence level of 95\%, 85\%, and 75\%, respectively.}
\label{intervals}
\end{figure}

Fig. \ref{intervals} visualizes the experimental results of 1-hour ahead interval prediction on the Milano dataset. When comparing ground truths (solid black line) with predictions (red dotted line), it can be observed that MetaSTNet can effectively capture the dynamic changes in traffic data. In addition, it shows how the prediction interval covers the ground truth at 95\%, 85\%, and 75\% confidence levels, respectively. It can also be observed that the vast majority of ground truths fall within the prediction interval at a 95\% confidence level.

\subsection{Scalability Analysis}
\begin{table}[t]
\centering
\caption{Complexity analysis for different models}
\label{tab:complexity}
\renewcommand{\arraystretch}{1.5}
\setlength{\tabcolsep}{8mm}{
\begin{tabular}{c|c}
\hline 
Methods   & Computational complexity   \\ \hline
DeepAuto  & $O(n^2d)$ \\
dmTP      & $O(n^2d)$    \\
STCNet    & $O(n^2d+nd^2)$  \\
ST-Tran   &  $O(n^2d+nd^2)$   \\ \hline
MetaSTNet &  $O(n^2d+nd^2)$   \\ \hline
\end{tabular}}
\end{table}

In this section, we have theoretically analyzed the computational complexity of MetaSTNet and other approaches to support scalability analysis. As DeepAuto, dmTP, STCNet, and ST-Tran perform better than other baseline models in our study, we would like to compare them with MetaSTNet. In particular, let $n,d$ denote the number and dimension of traffic data, respectively. The computational complexity of these five methods is shown in Table \ref{tab:complexity}. It can be observed that DeepAuto and dmTP show an exponential increase with the number of data and a linear increase with the dimension of data. However, DeepAuto and dmTP cannot capture the spatial dependencies of traffic data, resulting in poor prediction accuracy. Moreover, the computational complexity of STCNet, ST-Tran, and MetaSTNet are the same, which shows an exponential increase with the number and dimension of data. From this perspective, for the same computational complexity, our proposed model can obtain higher prediction accuracy than STCNet and ST-Tran using a small number of real-world training samples. Furthermore, it is seen that MetaSTNet can also assess the confidence of predicted volumes, which has a wider application in realistic scenarios.

\subsection{Efficiency Analysis}

\begin{figure}[!t]			
\centering{\includegraphics[width=\columnwidth]{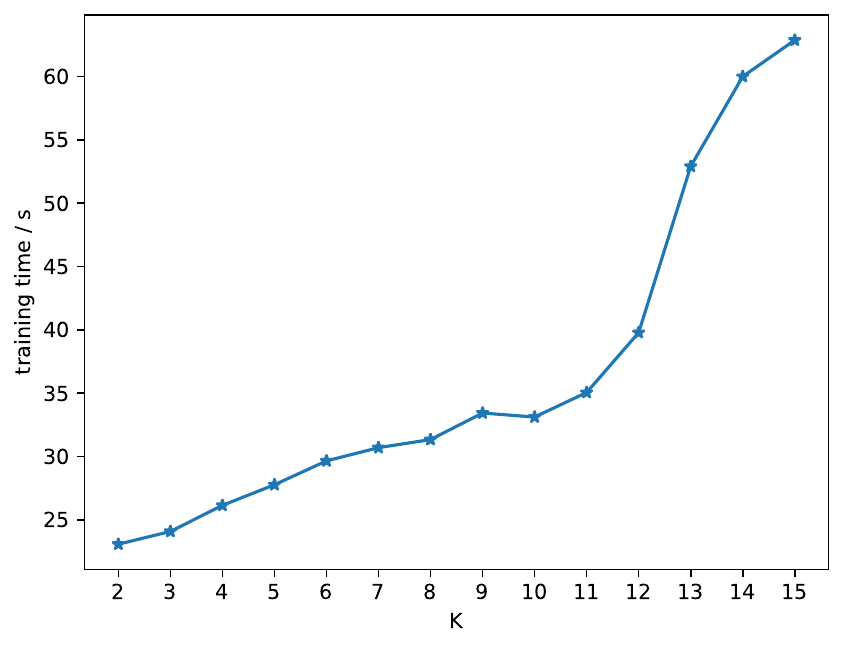}}
\caption{The training time with different $K$ on Milano dataset.} 
\label{training_time}		
\end{figure}

Fig. \ref{training_time} shows the training time of cross conformal prediction in MetaSTNet for different $K$ on the Milano dataset. It can be observed that when the number of $K$ is less than eleven, the training time shows a slow growth trend with the increase of $K$. When $K$ exceeds eleven, the training time increases significantly, indicating that a very large $K$ will significantly reduce the training efficiency.
    
\section{Conclusion}
We propose a deep learning model, MetaSTNet, for cellular network traffic prediction with sim-to-real transfer techniques. When the number of real-world traffic data is limited, it is wise to efficiently utilize simulation data to obtain optimal prediction accuracy for deep learning models. Besides, we design two parallel encoder-decoder structures to capture the closeness and period spatiotemporal dependencies of multimodal data. In addition, we develop a growing-window forward-validation scheme for cross conformal prediction to estimate the prediction intervals for the limited number of traffic data. To the best of our knowledge, this is the first time that we could quantify the confidence of cellular traffic predictions when only a small amount of traffic data is available. Furthermore, we evaluate the performance of different models on three real-world datasets and validate the effectiveness of our proposed MetaSTNet.

In future work, we would like to explore other deep learning methods to characterize the spatiotemporal correlations and improve the prediction performance. Moreover, centralized cellular network traffic prediction techniques meet problems of communication latency and data privacy leakage, as they require the transmission of a large amount of traffic data. Therefore, we plan to explore network traffic prediction techniques based on a large-scale distributed federated learning framework.

\bibliographystyle{IEEEtran}
\bibliography{refs}

\begin{IEEEbiography}
[{\includegraphics[width=1in,height=1.25in,clip,keepaspectratio]{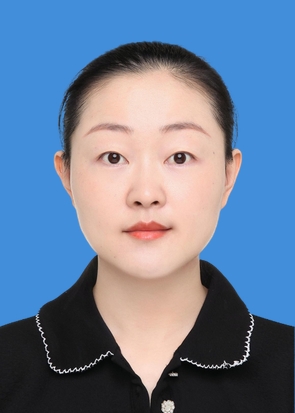}}]{Hui Ma}
received her B.Eng. degree in 2014 and M.S. degree in 2017 from Jiangnan University, Wuxi, China. She is a Ph.D. candidate in the Department of Computer Science, Tongji University, Shanghai, China. Her current research interests include deep learning and time series forecasting.
\end{IEEEbiography}

\begin{IEEEbiography}
[{\includegraphics[width=1in,height=1.25in,clip,keepaspectratio]{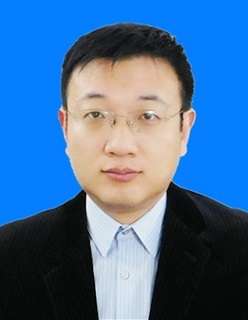}}]{Kai Yang}
received the B.Eng. degree from Southeast University, Nanjing, China, the M.S. degree from the National University of Singapore, Singapore, and the Ph.D. degree from Columbia University, New York, NY, USA.

He is a Distinguished Professor with Tongji University, Shanghai, China. He was a Technical Staff Member with Bell Laboratories, Murray Hill, NJ, USA. He has also been an Adjunct Faculty Member with Columbia University since 2011. He holds over 20 patents and has been published extensively in leading IEEE journals and conferences. His current research interests include big data analytics, machine learning, wireless communications, and signal processing.
\end{IEEEbiography}

\end{document}